# Agile Professional Virtual Community Inheritance via Adaptation of Social Protocols

## Willy Picard

Department of Information Technology
The Poznań University of Economics
ul. Mansfelda 4
60-854 Poznań, Poland
Email: picard@kti.ae.poznan.pl

**Abstract:** Support for human-to-human interactions over a network is still insufficient, particularly for professional virtual communities (PVC). Among other limitations, adaptation and learning-by-experience capabilities of humans are not taken into account in existing models for collaboration processes in PVC. This paper presents a model for adaptive human collaboration. A key element of this model is the use of negotiation for adaptation of social protocols modelling processes. A second contribution is the proposition of various adaptation propagation strategies as means for continuous management of the PVC inheritance.

## 1. Introduction

Enterprises are constantly increasing their efforts in order to improve their business processes. A main reason for this may be the fact that enterprises are exposed to a highly competitive global market. Among the most visible actions associated with this effort towards better support for better business processes, one may distinguish the current research work concerning Web services and associated standards: high-level languages such as BPEL (Alves et al., 2005) or WS-Coordination (Feingold and Jeyaraman, 2007) take the service concept one step further by providing a method of defining and supporting workflows and business processes.

However, it should be noticed that most of these actions are directed towards interoperable machine-to-machine interactions over a network. Support for *human-to-* *human* interactions over a network is still insufficient and more research has to be done to provide both theoretical and practical knowledge to this field.

Among various reasons for the weak support for human-to-human interactions, one may distinguish the following three reasons: first, many *social elements* are involved in the interaction among humans. An example of such a social element may be the role played by humans during their interactions. Social elements are usually difficult to model, e.g. integrating non-verbal communication to collaboration models. Therefore, their integration to a model of interaction between humans is not easy. A second reason is the *adaptation capabilities* of humans which are not only far more advanced than adaptation capabilities of software entities, but also not taken into account in existing models for collaboration processes. A third reason is the *learning-by-experience* capabilities of humans, i.e. the capabilities to extract know-how and knowledge from previous experience and reuse it in similar situations.

Human-to-human interactions between people sharing a common practice have been studied for many years. Wenger coined the term of Community of Practice (CoP) to refer to "a set of interacting people engaged in a common practice. Practice refers to the work people do, but also to the ideas behind it – the shared understandings and the activities." (Wenger 1998). More recently, Wenger, McDermott and Snyder (2002) refine the concept of CoP by proposing the following definition: "a set of people who share a concerns, a set of problems, or a passion about the topic, who deepen their knowledge and expertise in this area by interacting on an ongoing basis". Further refinements may be found in (Coakes and Clarke, 2006).

The concept of *professional virtual communities (PVCs)* has been proposed by the ECOLEAD project (2004-2008) and formalized by Bifulco and Santoro (20050 as a generalization of CoPs. While the studies on CoPs focus mainly on the interactions, and more specifically the "common practice", the interactions in PVCs may be classified in three areas: social, business and knowledge. While the core component of CoPs is the exchange of knowledge and experience via a common practice, social, business and knowledge elements are necessary for sustainable, motivated and durable PVCs (Crave and Vorobey, 2008).

The insufficient support for human-to-human interactions over a network is a strong limitation for a wide adoption of *professional virtual communities (PVCs)*. As mentioned in (Camarinha-Matos et al., 2005), "professional virtual community represents the combination of concepts of virtual community and professional community. Virtual communities are defined as social systems of networks of individuals, who use computer technologies to mediate their relationships. Professional communities provide environments for professionals to share the body of knowledge of their professions […]". According to Chituc and Azevedo (2005), little attention has been paid to the social perspective on Collaborative Networks (CN) business environment, including professional virtual communities in which social aspects are of high importance.

This paper is an attempt to provide a model for human-to-human interactions within professional virtual communities. The proposed model addresses, at least to some extent, the three characteristics of the interactions between humans. It should however been kept in mind that the results presented here are a work in progress and therefore they are not claimed to be neither sufficient nor exhaustive.

The rest of this paper is organized as follows. In section 2, the concept of *social protocol*, used to model collaboration processes, is presented. Section 3 then expands on *adaptation* of social protocols. Next, *agile PVC-inheritance* based on adaptation propagation strategies is discussed. Finally, section 5 concludes this paper.

## 2. Structuring Collaboration in PVCs

Appropriate support for structured collaboration in PVCs implies an analysis of PVCs as a sociosystem. Based on the characteristics of PVCs identified by such an analysis, an appropriate model of group interactions can be designed.

### 2.1 PVCs as Heterogeneous and Dynamic Environments

As defined by Ekholm and Fridqvist (1996), "a human *sociosystem* has a composition of human individuals, its structure is the social behaviour repertoire, i.e. interaction among human individuals". The sociosystem of professional virtual communities is highly *heterogeneous* and *dynamic*.

The heterogeneity of PVCs exists at various levels of granularity within PVCs. At a high level, a PVC consists usually of many different virtual teams (VTs). Each VT is different from other coexisting in the same PVC VTs in terms of goals, intentions, knowledge, processes, members, etc. At a lower level, one may notice that the structure of a VT is usually complex and heterogeneous. The roles played by the VT members, their skills, their competences are usually presenting a high level of diversity. A formal definition of VTs may be found in (Santoro and Bifulco, 2008).

Similarly to the heterogeneity of PVCs, the dynamics of PVCs exists at various levels of granularity within PVCs. At a high level, the set of VTs that the PVC consists of evolves in time: new VTs are created to answer new needs and opportunities, unnecessary VTs are dissolved, existing VTs change as new members enter and leave the community, etc. The dynamics of PVCs may hardly, not to say cannot, be foreseen at design time, as changes of a given PVC are naturally related to changes in its business environment (which is usually not a deterministic system). At a lower level, the structure of a VT is evolving in time: some members may have job promotion, the skills of the members are usually evolving (improve) in time. Additionally, members of a given VT may face new situations implying the development of new solutions, new ways of collaboration, etc.

The solutions proposed in our former work for support for heterogeneity and dynamics of PVCs are summarized in Table 1. The heterogeneity of both PVCs and VTs is addressed by the concept of *social protocols*. Dynamics of PVCs are addressed by *group actions*, while dynamics of VTs are addressed by *adaptation of social protocols*. These three concepts will shortly be presented in the next sections.

**Table 1**  Support for heterogeneity and dynamics of PVCs

|      | *Heterogeneity*   | *Dynamics*    |
|------|-------------------|---------------|
| PVCs | Social Protocols  | Group actions |
| VTs  | Social Protocols  | Adaptation    |

*2.2 Modeling Group Interactions with Social Protocols*

Support for human-to-human collaboration in PVCs should take into account the characteristics of PVCs as sociosystems presented in the former subsection, i.e. heterogeneity and dynamics.

*2.2.1. Overview of Social Protocols*

A first model for group interactions within a PVC has been presented in (Picard, 2005). The proposed model is based on the concept of *social protocol*. Social protocols model collaboration at a group level. The interactions of collaborators are captured by social protocols. Interactions are strongly related to social aspects, such as the role played by collaborators. The proposed model integrates some of these social aspects, which may explain the choice of the term "social protocols". Heterogeneity of PVCs at the VT level is then at least partially addressed by the social protocol approach.

A social protocol aims at modelling a set of collaboration processes, in the same way as a class models a set of objects in object-oriented programming. In other words, a social protocol may be seen as a model which instances are collaboration processes. Within a given PVC, various social protocols may be used to control interactions within different sub- communities, addressing at least partially the high level heterogeneity of PVCs.

Formally, a *social protocol p* is a finite state machine consisting of $\{ S_p, S_p^{start}, S_p^{end}, T_p \}$ where $S_p$ is the set of states, $S_p^{start} \subset S$ is the set of starting states, $S_p^{end} \subset S$ is the set of ending states, $S_p^{start} \cap S_p^{end} = \varnothing$, $T_p$ is the set of transitions from states to states.

In a social protocol, collaborators – as a group –move from state to state via the transitions. A transition may be triggered only by a collaborator labelled with the appropriate role. A transition is associated with the execution of an action. Execution of an action means the execution of remote code. SOAP or CORBA are examples of technologies that may be used to such remote code executions. A formal definition of the proposed model has been already presented in (Picard, 2006a), while an algorithm for structural validation of social protocols has been presented in (Picard, 2007b).

Social protocol example

*2.2.2. Social protocol example*

The example of social protocol which is presented in this Section is oversimplified for readability reasons. Social protocols modelling real-world collaboration processes are usually much more complex.

The chosen collaboration process to be modelled as a social protocol may be described as follows: a set of users are collaborating on the establishment of a "FAQ" (Frequently Asked Questions). Some users only ask questions, while others, referred as "experts" may answer the questions. Other users, referred as "managers", may interrupt the work on the FAQ document. The work on the document may terminate either by a success (the document has been written and the manager estimates that its quality is good enough to be published) or by a failure (the users did not find any way to collaborate and the manager has estimated that the work on the FAQ should be interrupted).

A possible model of this collaboration process as a social protocol is presented in Figure 1. Five states are represented as circles. State *"Waiting for first question"* is a starting state; states *"Failed termination"* and *"Successful termination"* are ending states. Transitions are represented as arrows with an icon representing the associated role and a text for the associated action.

**Figure 1** Example of social protocol

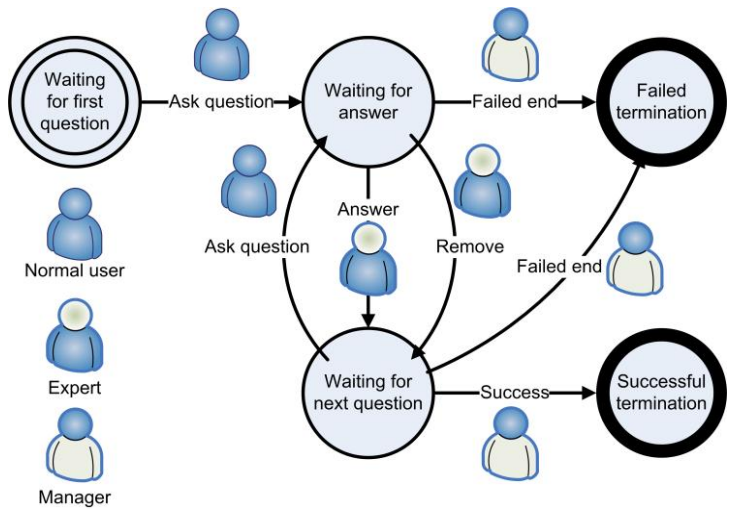

### 2.2.3. Group actions

A set of *group actions* have been identified to support *group dynamics*, i.e. the dynamics of PVCs at a high level. A group action is a special action that may be executed to modify the set of VTs that the PVC consists of. A group action may for instance allows a collaborator to split a group in two or more groups, or to merge two or more groups into a single group. Group dynamics may be modelled by a set of group actions. More details may be found in (Picard, 2005).

### 2.2.4. Abstract, (semi-) implemented social protocols and social processes

The concept of social protocol has been refined by introducing three types of social protocols: abstract, semi-implemented, and implemented.

An *abstract social protocol* is a definition of potential interactions among various abstract collaborators in an abstract environment. An abstract collaborator is an hypothetical human being, possessing given skills and playing a given social role. An example of an abstract collaborator may be a "logistics expert". The abstract environment refers to a set of potentially available services, without any related implementation. For instance, an abstract environment may possess message delivery means, whatever the implementation of this service may be (email, fax, or a message-oriented middleware). An abstract social protocol defines therefore collaboration in abstract means and requires additional specification of the implementation of both collaborators and actions.

An *implemented social protocol* is a definition of potential interactions among various identified collaborators, with a specification of all potential actions as service provided by the environment. In an implemented social protocol, all social roles are assigned to existing human beings, and potential actions may be executed by identified software entities. An implemented social protocol may therefore be instantiated as a *social process*.

A *social process* is an instantiation of an implemented social protocol. The state of the collaboration process (i.e. the current state) is stored in a social process which is ruled according to a given implemented social protocol. Using the same comparison as above, an abstract social protocol may be seen as an interface or abstract class, an implemented social protocol may be seen as an implemented class, while a social process may be seen as an object in the object-oriented programming paradigm.

Finally, *semi-implemented social protocols* are social protocols whose implementation is partially specified: some collaborators may already be identified, while some other collaborators still have to be identified. Similarly, the implementation of some actions may be known, while the implementation of other actions still have to be specified. The concept of semi-implemented social protocols is particularly important in the context of PVCs. Indeed, some recurrent services may be offered by the PVC. Therefore, some abstract social protocols may be semi-implemented with the help of services provided by the PVC, while other actions, depending on future VTs, may not be specified ex-ante.

Relations between abstract, semi-implemented, implemented social protocols and social processes are summarized in Table 2.

**Table 2**  Abstract, (semi-)implemented social protocols and social processes

|  | *Collaborators* | *Actions* | *Current state* | *Object-Oriented paradigm* |
|---|---|---|---|---|
| Abstract social protocol | Abstract | Abstract | N/A | Interface |
| Semi-implemented social protocol | Partially specified | Partially specified | N/A | Abstract class |
| Implemented social protocol | Fully specified | Fully specified | N/A | Class |
| Social process | Fully specified | Fully specified | Known | Object |

## 3. Adaptive Social Protocols

Social protocols address heterogeneity of PVCs at both high and low level, and dynamics at high level (with the help of group actions). However, the need for support for dynamics of PVCs is still only partially addressed at the VT level.

### 3.1 Run-time vs. Design-Time Adaptation

In the workflow management literature, information required to model and control a collaboration process has been classified according to various perspectives. In (van der Aalst et al., 2003), five perspectives have been presented:

- the *functional perspective* focuses on activities to be performed,

- the *process perspective* focuses on the execution conditions for activities,

- the *organization perspective* focuses on the organizational structure of the population that may potentially execute activities,
- the *information perspective* focuses on data flow among tasks,
- the *operation perspective* focuses on elementary operations performed by applications and resources.

A sixth perspective has been added in (Daoudi and Nurcan, 2003): the *intentional perspective* focuses on goals and strategies related to a given process. One may easily notice that all six perspectives presented above focus on elements that evolve in time.

In typical workflow management systems, two parts may be distinguished: a *design-time* part allows for definition of workflow schemas while the *run-time* part is responsible for execution of workflow instances. A main limitation of typical workflow management systems is the fact that once a workflow schema has been instantiated, the execution of the workflow instance must stick to the workflow schema till the end of the workflow instance execution. This limitation is not an issue if the lifespan of workflow instances is short in comparison with the time interval between two requests for changes of the workflow schema. When the lifespan of workflow instances is long in comparison with the time interval between two requests for changes of the workflow schema, a high number of workflow instances has to be executed with an ``incorrect'' workflow schema (i.e. that does not take into account required changes) or cancelled. As a consequence, typical workflow management systems are not flexible enough to support collaborative processes in two cases: highly dynamic, competitive markets/environments and long lasting collaboration processes.

In the case of highly dynamic, competitive markets/environments or long lasting collaboration processes, there is a strong need for the possibility to modify a workflow instance at run-time. Such modifications are usually needed to deal with situations which have not been foreseen nor modelled in the associated workflow schema. *Social protocol adaptation* refers to the possibility to *modify a running social protocol instance* to new situations which have not been foreseen and modelled in the associated social protocol.

### 3.2 Negotiation-based Adaptation

#### 3.2.1. Rationale for negotiation-base adaptation

While social protocols support, at least to some extent, the integration of some social elements (such as roles) to models of interactions among humans, the adaptation capabilities of humans are not taken into account into social protocols. There is however the need to provide adaptation mechanisms to social protocols. Indeed, interactions among humans are often a context-aware activity. In this paper, context-awareness refers to the capabilities of applications to provide relevant services to their users by sensing and exploring the users' context (Dey et al., 2001; Dockhorn et al., 2005). Context is defined as a "collection of interrelated conditions in which something exists or occurs" (Dockhorn et al., 2005). The users' context often consists of a collection of conditions, such as, e.g., the users' location, environmental aspects (temperature, light intensity, etc.) and activities (Chen et al., 2003). The users' context may change dynamically, and, therefore, a basic requirement for a context-aware system is its ability to sense context and to react to context changes.

In (Picard, 2006b), negotiations have been proposed as a method for adaptation of social protocols. The idea of negotiation of social protocol has been presented as "an attempt to weaken constraints usually limiting the interaction between collaborators, so that the adaptation capabilities of humans may be integrate in the life of a social protocol". The idea of using negotiations as an adaptation mean for social protocols comes from the fact that social protocols rule the interactions of all collaborators in a given group. Therefore each modification of the social protocol may influence all collaborators. As a consequence, the decision to modify a social protocol should be consulted and approved by many collaborators. Negotiations are a classical way to make collaborative decision and to reach an agreement in situations where expectations and goals of collaborators may be in conflict.

### 3.2.2. Layered adaptation

Adaptation of social protocols addresses changes in social processes, implemented and abstract social protocols. Indeed, when collaborators need to modify potential interactions in a given state of the social process, the result of the negotiation is a change of the implemented social protocol ruling the social process. As an implemented social protocol may be a particular "version" of an abstract social protocol, the modification of the implemented social protocol may lead to a modification of the associated abstract social protocol.

To illustrate the layered adaptation process, let's assume that a given group collaborates according to the abstract protocol presented in section 2.2.2. The abstract protocol needs to be implemented so that a social process may be instantiated. The following implementation is summarized in Tables 3 and 4.

**Table 3   Implementation of roles for the example social protocol**

| *Roles* | *Implementation* |
|---|---|
| Normal user | John Smith |
|  | Amy Tony |
| Expert | Bill Bogard |
|  | Jennifer Scott |
| Manager | Scott Tiger |
|  | Anna Gates |

**Table 4   Implementation of actions for the example social protocol**

| *Actions* | *Implementation (Web Services)* |
|---|---|
| Ask question | http://www.example.org/ws/askQuestion |
| Remove | http://www.example.org/ws/removeQuestion |
| Answer | http://www.example.org/ws/answerQuestion |
| Failed end | http://www.example.org/ws/suppressFAQ |
| Success | http://www.example.org/ws/publishFAQ |

During the collaboration process, after some questions have been asked and answered, Bill Bogard identifies that one answer formerly sent by Jennifer Scott should be commented. Currently the social protocol does not allow collaborators to interact in such a way. Then, Bill Bogard starts the process of adaptation of the social protocol, starting a negotiation process about the need for support for comments about answers. The chosen negotiation process concerns a relatively simple modification of the social protocol, i.e. the addition of a new transition from the state "Waiting for next question" to the same state, associated with the role "Expert" and implemented by the web service "http://www.example.org/ws/commentAnswer" provided by the environment of the group. During the negotiation process, Amy Tony suggests that normal user should also have the right to comment an answer, which is accepted by all the members of the group. As a consequence, the adaptation process leads to a new implemented social protocol, with two additional transitions (the first for the role "Expert", the second for "Normal User") from the state "Waiting for next question", associated with the web service formerly mentioned.

An abstract social protocol may be extracted from the adapted implemented social protocol, as presented in Figure 2. In this adapted abstract social protocol, the two newly proposed transitions have been added, but no implementation is proposed for the abstract action "comment".

**Figure 2**   Adapted abstract social protocol

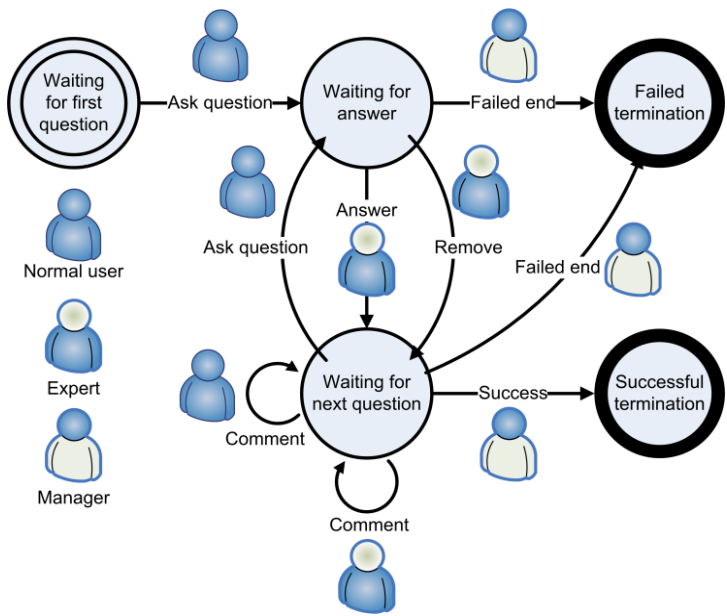

## 4. Adaptation of Social Protocols in PVCs

In the context of PVCs, adaptation leads to support for dynamics of collaboration processes at the group level. Additionally, decisions taken during adaptation of social protocols may be reused by other groups facing similar problems.

*4.1 Adaptation Propagation Strategies*

Adaptation of a social protocol in a given group leads to the creation of a new version of the social protocol ruling collaboration within this group. Let's assume that the adaptation of a given social protocol $P_1$ in a given group $G$ leads to the creation of a new social protocol $P_1'$. In the context of a PVC, various strategies may be used to manage the change caused by the adaptation of a social protocol:

- *Local adaptation strategy*: Other groups ruled by the social protocol $P$ are not affected by the adaptation and are still ruled by $P$. The social protocol $P'$ is only used by group $G$ and is not available for future groups.

- *Global propagation strategy*: Other groups ruled by the social protocol $P$ are not affected by the adaptation and are still ruled by $P$. The social protocol $P'$ is used by group $G$ and is available for future groups.

- *Instant propagation strategy*: Other groups ruled by the social protocol $P$ are affected by the adaptation, as they are now ruled by $P'$. The social protocol $P'$ replaces $P$ in the whole PVC.

It should be noticed that the instant propagation strategy may not always be used as the changes provided by the adaptation of the social protocol may be in conflict with the current state of some collaboration processes.

Additionally, adaptation propagation is not always possible because of difference in terms of available services in various environments. If two groups, working in two different environments in which the sets of available services are different, modifications provided by collaborators of one group may not always be propagated to the second group. For instance, let's assume that two groups $G_1$ and $G_2$ collaborate according to the implemented protocol presented in section 3.2.2. If the group $G_1$ adapts the social protocol as presented in section 3.2.2, i.e. adds two transitions so that experts and normal users may comment answers, then the abstract social protocol is modified. However, the group $G_2$ may take advantage of this adaptation iff the action "comment" may be implemented, i.e. it exists an implementation of this action in the environment of $G_2$.

While the layered adaptation may at the first sight seen as a limitation, it is a major improvement in the proposed adaptation mechanism. Indeed, in the case when the action implementation used by the group $G_1$ is not available to $G_2$, the second group has still the possibility to choose another implementation of the "comment" action. Therefore, the adaptation propagation may be now done at the abstract level, allowing various groups to take advantage of the changes proposed by other groups sharing the same abstract social protocol, but with an additional degree of freedom for the implementation of actions.

*4.2 Adaptation Propagation in a VO-Inheritance Management Perspective*

The concept of *virtual organization inheritance* (VO-I) has been defined in (Loss et al., 2006a) as "the set of information and knowledge accumulated from past and current VOs along their entire life cycle. *Virtual organization inheritance management* (VO-I-M) corresponds to the VO activity that manages what has been inherited about given VOs, usually supported by computer systems".

**Figure 3** VO-I a) before adaptation b) with a local adaptation strategy c) with a global propagation strategy d) with an instant propagation strategy

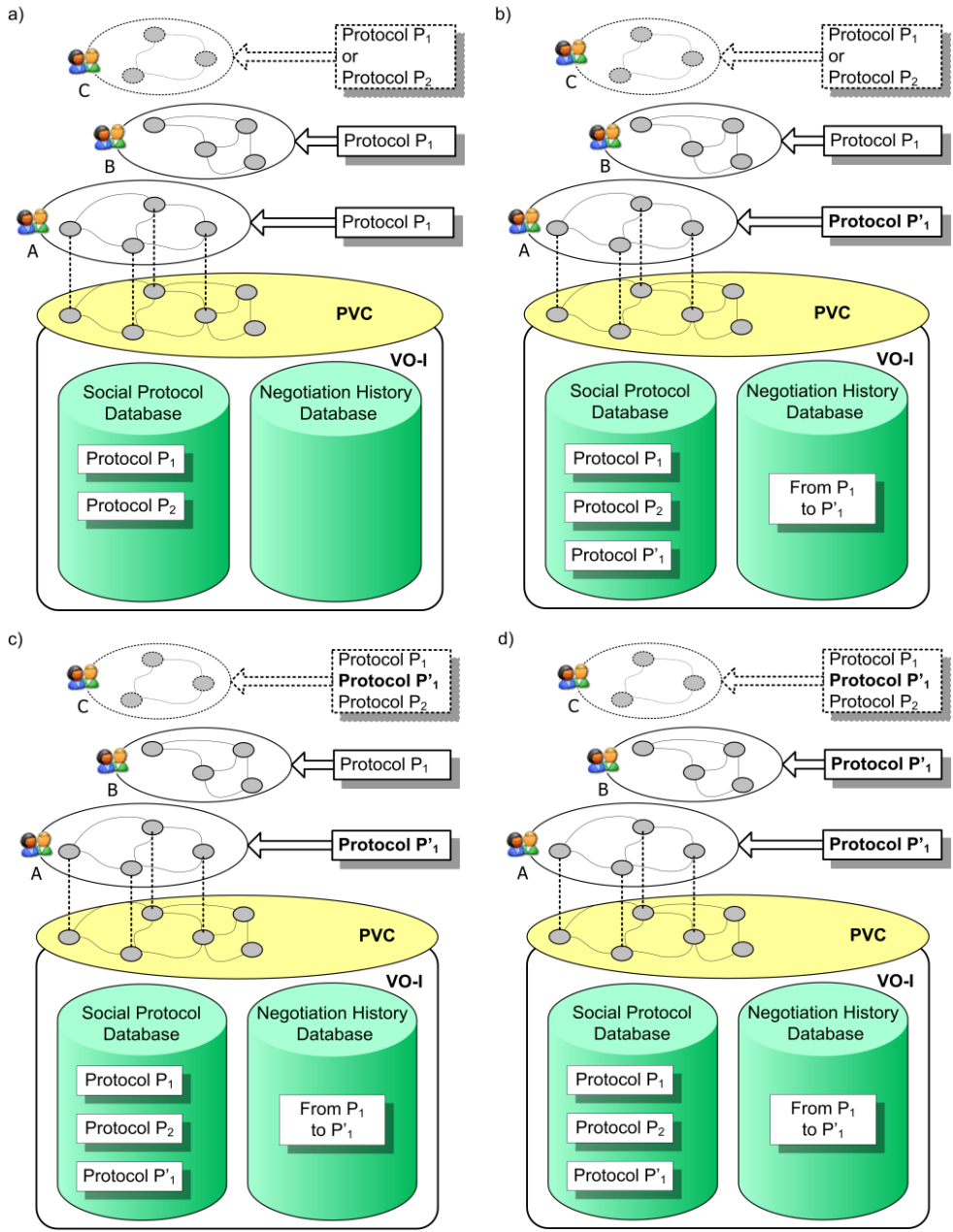

In a VO-I-M perspective, adaptation of social protocols may be seen as part of the VO-I, as presented in Figure 3. In the PVC presented in Figure 3a), two protocols are available – $P_1$ and $P_2$ – and two VTs – $A$ and $B$ – are ruled by $P_1$. A new VT $C$ may be created with either protocol $P_1$ or protocol $P_2$. It is then assumed that the VT $A$ has adapted the protocol $P_1$, which leads to the protocol $P'_1$. Figure 3b) (respectively 3c) and 3d)) illustrates the state of the PVCs after the adaptation in the case of a local adaptation strategy (respectively global and instant adaptation strategy). In Figure 3b), the newly created protocol $P'_1$ rules the VT *A* but is not available to VTs *B* and *C*. In Figure 3c), the newly created protocol is available to the new VT *C* but the VT *B* is still ruled by $P_1$. In Figure 3d), $P'_1$ is available to the new VT *C* and the VT *B* is now ruled by $P'_1$.

The newly created social protocol $P'_1$ embeds knowledge about an alternative way to collaborate. The social protocol $P'_1$ models the additional knowledge and expertise which have been required to react to situations which have not been unforeseen nor modelled in the social protocol $P_1$. Information about the negotiation process that leads from $P_1$ to $P'_1$ is available from the negotiation history database, as presented in Figure 3. One should notice that these knowledge and expertise should not necessarily been directly reused, but could be used for consultation about what has happened in similar cases and the solution found. Additionally, privacy should be taken into account. Collaborators that negotiated the social protocol $P_1$ should explicitly agree to publish negotiation-related information before such information is available to other VTs.

The global propagation strategy would allow collaborators of VTs to consult and eventually reuse the VO-I of VTs in which a social protocol has been adapted. The instant propagation strategy would enforce the reuse of newly-created knowledge by other VTs in a normative way: the adapted social protocol "overwrites" the original social protocol.

Finally, the proposed adaptation propagation strategies provide means for *continuous VO-I-M*, which leads to agile PVCs. A classical issue in VO-I-M is the frequency of VO-I capturing. A briefing-debriefing technique has been presented by Loss et al. (2006b), proposing to capture VO-I by comparing the results of two interview meetings: usually the first interview meeting takes place before the VO is created, while the second one (the debriefing) takes place after VO dissolution or metamorphosis. The briefing-debriefing technique may be used "to double-check the plans, fine tune the assignments of tasks, rehearsal the actions and also to exchange lessons learned, evaluate the actions against the plans and to register explicitly the knowledge acquired, respectively". Therefore, the briefing-debriefing technique may capture more elements of the VO-I, than just those related to social protocols. On the second hand, information about adaptation of a social protocol would be captured by the briefing-debriefing technique during the debriefing session, while the adaptation propagation strategies make information about adaptation of a social protocol accessible by other VOs just after the adaptation. Therefore, propagation strategies may enable continuous VO-I-M of social protocols, while the briefing-debriefing technique is less agile but may capture more elements of the VO-I.

## 5. Conclusions

The introduction of adaptation of social protocols and adaptation propagation strategies provides computer support to management of PVC-inheritance related to collaboration processes. To our best knowledge, it is the first attempt to support continuous management of VO-inheritance, even if the proposed solution is limited to PVC-inheritance elements related to collaboration processes.

The main contributions presented in this paper are 1) a layered approach to the concept of social protocols allowing separation of collaboration structure from implementation, 2) the rationale for adaptation of social protocols in PVCs as heterogeneous and dynamic sociosystems, 3) three strategies for adaptation propagation, 4) the proposition of adaptation of social protocols and adaptation propagation as means for continuous management of PVC inheritance.

The layered approach to social protocols and adaptation propagation are complementary, enabling a sound foundation for agile PVCs. PVCs supporting abstract social protocols and adaptation propagation would support virtual teams by, on the first hand, providing support for structured interactions among collaborators, on the second hand, allowing collaborators to modify social protocols ruling their interactions and sharing their experience with other virtual teams collaborating in a similar way (i.e. sharing the same social protocol).

In a broader perspective, the adaptation of social protocols and its potential propagation may lead to similar changes in the area of workflow support systems as we have witness with contents with the rise of the Web 2.0. Indeed, the adaptation of social protocols would blur the classical distinction between protocol "producers" (or process designer) and protocol "customers" (or process actors), as the Web 2.0 blurs the distinction between content producer and content consumer.

Among future works, a formal model of propagation strategies presented in this paper should be established and validated by experiments. A prototype is currently under implementation and will be tailored to the needs of a pilot for the construction sector. In the planned pilot for the construction sector, the solution presented in this paper has to be refined to support Virtual Breeding Environments(VBEs), and not only PVCs. The main challenge for the application of the presented solution to VBEs is the fact that members of virtual organizations (VOs) are non-monolithic, i.e. each member of a VO consists of many individuals with various skills, culture, goals, social networks, etc. Therefore, the concepts and models presented in this paper have to be adapted to support the duality of human-to-human interactions in VBEs: interactions in VBEs occur among humans as individuals, as well as among humans as members of an organization participating in VOs.

## Remark

This paper is an extended version of the paper entitled "Continuous Management of Professional Virtual Community Inheritance Based on the Adaptation of Social Protocols" published in "Establishing the foundation of Collaborative Networks, Proc. Of the 8th IFIP Working Conference on Virtual Entreprises" (Picard, 2007a).

## References

Alves, A., Arkin, A., Askary, S., Barreto, C., Bloch, B., Curbera, F., Ford, M., Goland, Y., Guízar, A., Kartha, M., Liu, C.K., Khalaf, R., König, D., Marin, M., Mehta, V., Thatte, S., van der Rijn, D., Yendluri, P. and Yiu, A., editors. (2007) *Web Services Business Process Execution Language Version 2.0*. WS-BPEL TC OASIS, April 2007.
Available via http://docs.oasis-open.org/wsbpel/2.0/OS/wsbpel-v2.0-OS.pdf

van der Aalst, W.M.P., Weske, M. and Wirtz, G. (2003) Advanced topics in workflow management: Issues, requirements, and solutions. *J. of Integrated Design and Process Science*, **7**(3), pp. 49–77

Bifulco, A. and Santoro, R. (2005) A Conceptual Framework for "Professional Virtual Communities". In *Collaborative Networks and their Breeding Environments*, Proc. of the 6th IFIP Working Conf. on Virtual Enterprises (PRO-VE 2005), Valencia, Spain, Sept. 26-28, 2005, Springer, pp. 417–424

Camarinha-Matos, L.M., Afsarmanesh, H. and Ollus, M. (2005) ECOLEAD: A Holistic Approach to Creation and Management of Dynamic Virtual Organizations. In *Collaborative Networks and their Breeding Environments*, Proc. of the 6th IFIP Working Conf. on Virtual Enterprises (PRO-VE 2005), Valencia, Spain, Sept. 26-28, 2005, Springer, pp. 3–16

Chen, H., Finin, T. and Joshi, A. (2003) An Ontology for Context-Aware Pervasive Computing Environments. *Knowledge Engineering Review, Special Issue on Ontologies for Distributed Systems*, **18**(3) Cambridge University Press, pp. 197–207

Chituc, C.M., and Azevedo, A.L. (2005) Multi-Perspective Challenges on Collaborative Networks Business Environments. In *Collaborative Networks and their Breeding Environments*, Proceedings of the 6th IFIP Working Conf. on Virtual Enterprises (PRO-VE 2005), Valencia, Spain, Sept. 26-28, 2005, Springer, pp. 25–32

Coakes, E. and Clarke, S. (2006) The Concept of Communities of Practice. In *Encyclopedia of Communities of Practice in Information and Knowledge Management*. 2006, Idea Group Inc.

Crave, S. and Vorobey, V. (2008) Business Models for PVC: Challenges and Perspectives. In *Methods and Tools for Collaborative Networked Organizations*, Camarinha-Matos, L.M., Afsarmanesh, H. and Ollus, M. (eds), 2008, Springer, pp. 295–306

Daoudi, F. and Nurcan, S. (2007) A benchmarking framework for methods to design flexible business processes. *Special Issue on Design for Flexibility of the "Software Process: Improvement and Practice Journal"*, 12(1), pp. 51–63

Dey, A.K., Salber, D. and Abowd, G. D. (2001) A Conceptual Framework and a Toolkit for Supporting the Rapid Prototyping of Context-Aware Applications. *Human-Computer Interaction*, **16**(2-4), pp. 97–166

Dockhorn Costa, P., Ferreira Pires, L. and van Sinderen, M. (2005) Designing a Configurable Services Platform for Mobile Context-Aware Applications. *Int. J. of Pervasive Computing and Communications (JPCC)*, **1**(1), pp. 27–37

ECOLEAD project (2004-2008). European collaborative networked organizations leadership initiative. FP6-IST-506958. Homepage: http://ecolead.vtt.fi/

Ekholm, A. and Fridqvist, S. (1996) Modelling of user organisations, buildings and spaces for the design process. In *Construction on the Information Highway*, Proceedings from the CIB W78 Workshop, Bled, Slovenia, June 10-12, 1996

Feingold, M. and Jeyaraman, R. (2007) *Web Services Coordination (WS-Coordination) Version 1.1.* OASIS Web Services Transaction WS-TX TC, 12 July 2007. Available via http://docs.oasis-open.org/ws-tx/wstx-wscoor-1.1-spec.pdf

Loss, L., Pereira-Klen, A.A. and Rabelo, R.J. (2006a) Knowledge Management Based Approach for Virtual Organization Inheritance. In *Network-centric Collaboration and Supporting Frameworks*, Proc. of the 7th IFIP Working Conf. on Virtual Enterprises (PRO-VE 2006), Helsinki, Finland, Sept. 2006. Springer, pp. 285–294

Loss, L., Rabelo, R.J. and Pereira-Klen, A.A. (2006b) Virtual Organization Management: An Approach Based on Inheritance Information. In: *Global Conference on Sustainable Product Development and Life Cycle Engineering*, São Carlos, SP, Brazil, Editora Suprema

Picard, W. (2007a) Continuous Management of Professional Virtual Community Inheritance Based on the Adaptation of Social Protocols. In , *Establishing the foundation of Collaborative Networks*, Proc. Of the 8th IFIP Working Conference on Virtual Entreprises (PRO-VE 2007), Guimaraes, Portugal, Sept. 2008, pp. 381–388.

Picard, W. (2007b) An Algebraic Algorithm for Structural Validation of Social Protocols. In *Lecture Notes in Computer Science*, 4439, Springer, pp. 570–583

Picard, W. (2006a) Adaptive Collaboration in Professional Virtual Communities via Negotiations of Social Protocols. In *Network-centric Collaboration and Supporting Frameworks*, Proc. of the 7th IFIP Working Conf. on Virtual Enterprises (PRO-VE 2006), Helsinki, Finland, Sept. 2006. Springer, pp. 353–360

Picard, W. (2006b) Adaptive Human-to-Human Collaboration via Negotiations of Social Protocols". In *Technologies for Business Information Systems*, Proc. of the 9th Int. Conf. on Business Information Systems, Klagenfurt, Austria, May 31 – June 2, 2006, Springer Verlag, pp. 193–203

Picard, W. (2005) Modeling Structured Non-monolithic Collaboration Processes. In *Collaborative Networks and their Breeding Environments*, Proc. of the 6th IFIP Working Conf. on Virtual Enterprises (PRO-VE 2005), Valencia, Spain, Sept. 26-28, 2005, Springer, pp. 379–386

Santoro, R. and Bifulco, A. (2008) Professional Virtual Communities Reference Framework. In *Methods and Tools for Collaborative Networked Organizations*, Camarinha-Matos, L.M., Afsarmanesh, H. and Ollus, M. (eds), 2008, Springer, pp. 277–294

Wenger, E., McDermott, R. and Snyder, W.M. (2002) *Cultivating Communities of Practice: A Guide to Managing Knowledge*, Boston, USA. Harvard Business School Press.

Wenger, E. (1998) *Communities of Practice: Learning, Meaning and Identity.* Cambridge, UK. Cambridge University Press